\documentclass[12pt,a4paper]{article}
\usepackage{amssymb}
\usepackage{graphicx,subfigure}
\def\beq{\begin{equation}}
\def\eeq{\end{equation}}
\def\beqr{\begin{eqnarray}}
\def\eeqr{\end{eqnarray}}
\def\bdpm{\begin{displaymath}}
\def\edpm{\end{displaymath}}
\def\nuc#1#2#3 {Nucl. Phys. {\bf#1}, #2 (#3)}
\def\mpla#1#2#3 {Mod. Phys. Lett. A {\bf#1}, #2 (#3)}
\def\plb#1#2#3 {Phys. Lett. B {\bf#1}, #2 (#3)}
\def\prd#1#2#3 {Phys. Rev. D {\bf#1}, #2 (#3)}
\def\prl#1#2#3 {Phys. Rev. Lett. {\bf#1}, #2 (#3)}
\def\ptp#1#2#3 {Prog. Theor. Phys. {\bf#1}, #2 (#3)}
\def\rmp#1#2#3 {Rev. Mod. Phys. {\bf#1}, #2 (#3)}
\def\zpc#1#2#3 {Z. Phys. C {\bf#1}, #2 (#3)}
\def\ibid#1#2#3 {{\it ibid.} {\bf#1}, #2 (#3)}
\def\none#1#2#3 {{\bf#1}, #2 (#3)}
\newcommand{\CP}{\textit{CP}}
\newcommand{\CKM}{\textbf{CKM}}
\newcommand{\SM}{\textbf{SM}}
\newcommand{\NP}{\textbf{NP}}
\newcommand{\LRM}{\textbf{LRM}}
\newcommand{\OPE}{\textbf{OPE}}
\newcommand{\QCD}{\textbf{QCD}}

\newcommand{\LL}{\textbf{LL}}

\begin{document}
\begin{titlepage}
\begin{flushright}
UMN-TH-2211/03 \\
FTPI-MINN-03/22 \\
August 2003
\end{flushright}
\vspace{0.7in}
\begin{center}
{\Large \bf $CP$ asymmetries in penguin-induced B decays
  in general left-right models \\ }
\vspace{1.2in} {\bf Soo-hyeon Nam \\ }
William I. Fine Theoretical Physics Institute, University of Minnesota, \\
Minneapolis, MN 55455, USA \\
\vspace{1.2in} {\bf Abstract \\ }
\end{center}
We study \CP\ asymmetries in penguin-induced $b \rightarrow s\bar{s}s$ decays in general
left-right models without imposing manifest or pseudomanifest left-right symmetry.
Using the effective Hamiltonian approach, we evaluate \CP\ asymmetries
in $B^\pm \rightarrow \phi K^{(\ast)\pm}$ decays as well as mixing induced $B$ meson decays
$B \rightarrow J/\psi K_s$ and $B \rightarrow \phi K_s$ decays.
Based on recent measurements revealing large \CP\ violation, we show that nonmanifest type
model is more favored than manifest or pseudomanifest type.
\end{titlepage}

\section{Introduction}
 One of the major goals of present experiments in $B$ physics is the study of \CP\ violation
which may reside in the quark flavor mixing described by the Cabibbo-Kobayashi-Maskawa
(\CKM) matrix in the Standard $SU(2)_L\times U(1)$ Model (\SM).  Since there is one complex phase
in \CKM\ matrix, the sizes and patterns of \CP\ violation in various decay modes in the \SM\ are
in principle expressed through this single parameter \cite{Review}.
But the present experimental results with large \CP\ violation effects in the $B$ meson system
are not simply explained with this single parameter under the minimal \SM\ framework \cite{Nir02}.
For instance, the \CP\ asymmetries in mixing induced $B$ meson decays
is characterized by a \CP\ angle $\beta$ which is a phase of the \CKM\ matrix element $V_{td}$, and
the observed world average value of $\sin2\beta$ in $B \rightarrow J/\psi K_S$
($b \rightarrow c\bar{c}s$) decays is given by
\beq
\sin2\beta_{J/\psi K_S} = 0.734 \pm 0.054 .
\eeq
Besides, this \CP\ angle $\beta$ is recently measured by \textbf{BABAR} and \textbf{Belle}
in $B \rightarrow \phi K_S$ ($b \rightarrow s\bar{s}s$) decays \cite{exp1}, and their average
value is
\beq
\sin2\beta_{\phi K_S} =  - 0.39 \pm 0.41 .
\eeq
In the \SM , however, the \CP\ asymmetry in $B \rightarrow \phi K_S$ decays is
expected to be very close to that in $B \rightarrow J/\psi K_S$ decays \cite{Grossman97}.
Admitting that the statistical error of those experimental data is still
large to confirm the data and justify any theory, a $2.7\sigma$ deviation between
$\sin2\beta_{J/\psi K_S}$ and $\sin2\beta_{\phi K_S}$ may give a clue of new physics (\NP) effects
in $B$ decays. If so, other inclusive $b \rightarrow s\bar{s}s$ dominated $B$ decays such as
$B^\pm \rightarrow \phi K^{(\ast)\pm}$ decays might receive the same contribution
from the \NP .

  In a recent paper \cite{Nam02}, we have investigated the mixing induced \CP\ asymmetry
in $B \rightarrow J/\psi K_S$ decays in the general left-right model (\LRM) with group
$SU(2)_L\times SU(2)_R\times U(1)$ since it is one of the simplest extensions of the \SM\ gauge
group as a complement of the purely left-handed nature of the \SM\ \cite{Pati74}.
Due to the extended group $SU(2)_R$ in the \LRM\, there are new neutral and charged gauge
bosons, $Z_R$ and $W_R$ as well as a right-handed gauge coupling, $g_R$.
After spontaneous symmetry breaking, the gauge eigenstates $W_R$ mix with $W_L$ to form
the mass eigenstates $W$ and $W^\prime$ with masses $M_W$ and $M_{W^\prime}$, respectively.
The $W_L-W_R$ mixing angle $\xi$ and the ratio $\zeta$ of $M_W^2$ to $M_{W^\prime}^2$ are
restricted by a number of low-energy phenomenological constraints along
with the right-handed mass mixing matrix elements. From the limits on deviations of
muon decay parameters from the V-A prediction, the lower bound on $M_{W^\prime}$ can be
obtained as follows \cite{Balke88}:
\beq
\zeta_g < 0.033 \qquad \textrm{or} \qquad M_{W^\prime} > (g_R/g_L)
 \times 440\ \textrm{GeV} ,
\label{MWRbound}
\eeq
where $\zeta_g \equiv g_R^2 M_W^2 / g_L^2 M_{W^\prime}^2$.
Previously, stronger limits of the mass $M_{W^\prime}$ as well as the mixing angle $\xi$
were presented by many authors experimentally \cite{exp2} and theoretically \cite{Lang89}
assuming manifest ($V^R=V^L$) or pseudomanifest ($V^R=V^{L\ast}K$) left-right symmetry ($g_L=g_R$),
where $V^L$ and $V^R$ are the left- and right-handed quark mixing matrices, respectively, and
$K$ is a diagonal phase matrix \cite{Moha92}.  But, in general, the form of $V^R$ is not
necessarily restricted to manifest or pseudomanifest symmetric types, so the $W_R$ mass limit
can be lowered to approximately 300 GeV by taking the following forms of $V^R$ \cite{Olness84}:
\beq
V^R_I = \left( \begin{array}{ccc} e^{i\omega} & \sim 0 & \sim 0 \\
                            \sim 0 & c_R e^{i\alpha_1} & s_R e^{i\alpha_2} \\
                       \sim 0 & -s_R e^{i\alpha_3} & c_R e^{i\alpha_4} \end{array} \right) ,\quad
V^R_{II} = \left( \begin{array}{ccc} \sim 0 & e^{i\omega} & \sim 0 \\
                            c_R e^{i\alpha_1} & \sim 0 & s_R e^{i\alpha_2} \\
                            -s_R e^{i\alpha_3} & \sim 0 & c_R e^{i\alpha_4} \end{array} \right) ,
\label{VR}
\eeq
where $c_R\ (s_R)\equiv \cos\theta_R\ (\sin\theta_R)$ $(0^\circ \leq \theta_R \leq 90^\circ )$.
Here the matrix elements indicated as $\sim 0$ may be $\lesssim 10^{-2}$ and unitarity
requires $\alpha_1+\alpha_4=\alpha_2+\alpha_3$. From the $b\rightarrow c$ semileptonic
decays of the $B$ mesons, we can get an approximate bound
$\xi_g\sin\theta_R \lesssim 0.013$ by assuming $|V^L_{cb}|\approx 0.04$ \cite{Voloshin97}
, where $\xi_g \equiv (g_R/g_L)\xi$.\footnote{In Ref. \cite{Nam02}, $\xi_g$ is defined as
$(g_L/g_R)\xi$ unlike this paper so that the mistakenly written bound
$\xi_g\sin\theta_R \lesssim 0.013$ should read $(g_R/g_L)\xi\sin\theta_R \lesssim 0.013$.}
This new parameter $\xi_g$ is in general smaller than
the charged gauge boson mass ratio $\zeta_g$ in the general \LRM\ \cite{Nam02,Lang89}.
In a similar way to the charged gauge bosons, the neutral gauge bosons mix each other
\cite{Chay}.  But we do not present them here because $Z_R$ contribution to penguin-induced
B decays is negligible. Also, due to gauge invariance, tree-level flavor-changing neutral Higgs
bosons with masses $M_H$ enter into our theory \cite{Chang84}.  However, we also neglect
their contributions by assuming $M_H \gg M_{W^\prime}$.

  The \CP\ asymmetry in the penguin-induced $B \rightarrow \phi K_S$ decays was also studied
earlier in the pseudomanifest left-right symmetry model in Ref. \cite{Baren98}.  In this case,
the right-handed current contribution to $B\bar{B}$ mixing is suppressed by $\zeta$
so that the \NP\ effect only arises in the magnetic penguin since the suppression by $\xi$
is offset by a large factor $m_t/m_b$ arising in the virtual top quark loop \cite{Cho94}.
However, in the nonmanifest \LRM , $\zeta$ terms in $B\bar{B}$ mixing and absorptive part of
the decay amplitudes become important due to the possible enhancement of $V^R$ elements
so that the right-handed current contribution to the corresponding \CP\ asymmetry is more enhanced.
In this paper, as a continuation of our previous work, we will explicitly evaluate
the possible right-handed current contribution to \CP\ asymmetry
in $B^\pm \rightarrow \phi K^{(\ast)\pm}$ decays as well as in $B \rightarrow \phi K_S$ decays
in the general \LRM\ related to recent measurements, and show that \CP\ asymmetries in those decays
can be large enough to probe the existence of the right-handed current using the effective
Hamiltonian approach.  After reviewing the structure of the effective Hamiltonian in the general
\LRM\ in Sec. 2 , we will discuss \CP\ asymmetries in the several $b \rightarrow s\bar{s}s$
dominated $B$ decays in Sec. 3 in detail.

\section{Effective Hamiltonian}

 The low-energy effects of the full theory can be described by the effective Hamiltonian
approach in order to include \QCD\ effects systematically.  The low-energy
effective Hamiltonian calculated within the framework of the operator product expansion (\OPE)
has a finite number of operators in a given order, which is dependent upon the structure of the
model.  In the \LRM , the low energy effective Hamiltonian at the energy scale $\mu$
for $\Delta B = 1$ and $\Delta S = 1$ transition has the following form:
\beqr
\mathcal{H}_{eff} &=& \frac{G_F}{\sqrt{2}} \Big[
  \sum_{{i=1,2}\atop{q=u,c}}\lambda_q^{LL} C_i^q O_i^q  - \lambda_t^{LL}
  \big(\sum_{i=3}^{12} C_i O_i  + C_7^\gamma O_7^\gamma + C_8^G O_8^G\big) \Big] \nonumber \\
  &+& (C_i O_i \rightarrow C'_i O'_i) ,
\eeqr
where $\lambda_q^{AB}\equiv V^{A\ast}_{qs}V^B_{qb}$,
$O_{1,2}$ are the standard current-current operators, $O_3 - O_{10}$ are
the standard penguin operators, and $O_7^\gamma$ and $O_8^G$ are the standard photonic and
gluonic magnetic operators, respectively, which can be found in Ref. \cite{Buch96}.  Since we have
additional $SU(2)_R$ group in the \LRM , the operator basis is doubled by $O'_i$ which
are the chiral conjugates of $O_i$.  Also new operators $O_{11,12}$ and $O'_{11,12}$
arise with mixed chiral structure of $O_{1,2}$ and $O^\prime_{1,2}$ \cite{Cho94}.

 In order to calculate the Wilson coefficients $C_i(\mu)$, we first calculate them at $\mu = M_W$
scale.  After performing a straightforward matching computation, we find the Wilson coefficients
at $W$ scale neglecting the $u$-quark mass:
\beqr \label{CoefMW}
C^q_2(M_W) &=& 1 , \quad
C^{q\prime}_2(M_W)\ =\ \zeta_g\lambda^{RR}_q/\lambda^{LL}_q , \cr
C_7^\gamma(M_W) &=& F(x_t^2) + A^{tb}\widetilde{F}(x_t^2) , \cr
C_7^{\gamma\prime}(M_W) &=& A^{ts\ast}\widetilde{F}(x_t^2) , \\
C_8^G(M_W) &=& G(x_t^2) + A^{tb}\widetilde{G}(x_t^2) , \cr
C_8^{G\prime}(M_W) &=& A^{ts\ast}\widetilde{G}(x_t^2) , \nonumber
\eeqr
where
\beq
x_q = \frac{m_q}{M_W}\ (q=u,c,t), \qquad
A^{tD} = \xi_g \frac{m_t}{m_b}\frac{V^R_{tD}}{V^L_{tD}}e^{i\alpha_\circ} \ (D=b,s),
\eeq
and $\alpha_\circ$ is a \CP\ phase residing in the vacuum expectation values, which can be absorbed
in $\alpha_i$ in Eq. (\ref{VR}) by redefining $\alpha_i + \alpha_\circ \rightarrow \alpha_i$.
All other coefficients vanish.  In Eq. (\ref{CoefMW}), the explicit forms of the functions
$F(x_t)$, $\widetilde{F}(x_t)$, $G(x_t)$, and $\widetilde{G}(x_t)$ are given in Ref. \cite{Cho94},
and the terms proportional to $\xi_g$ and $\zeta_g$ in the magnetic coefficients are neglected
except the contribution coming from the virtual $t$-quark which gives $m_t/m_b$ enhancement.
Also the term proportional to $\zeta_g$ in the tree-level coefficient $C'_2$ is not neglected
because $\zeta_g \geq \xi_g$ and there is possible enhancement by the ratio of
\CKM\ angles ($\lambda^{RR}_q/\lambda^{LL}_q$) in the nonmanifest \LRM .

  The coefficients $C_i(\mu)$ at the scale $\mu = m_b$ can be obtained by evolving the coefficients
$C_i(M_W)$ with the $28\times 28$ anomalous dimension matrix applying the usual renormalization
group procedure.  Since the strong interaction preserves chirality, the $28\times 28$ anomalous
dimensional matrix decomposes into two identical $14\times 14$ blocks.  The \SM\ $12\times 12$
submatrix describing the mixing among $O_1 - O_{10}$, $O_7^\gamma$, and $O_8^G$ can be found
in Ref. \cite{Ciuchini93}, and the explicit form of the remaining $4\times 4$ matrix describing
the mixing among $O_{11,12}$, $O_7^\gamma$, and $O_8^G$, which partially overlaps with the
\SM\ $12\times 12$ submatrix, can be found in Ref. \cite{Cho94}.  The low energy Wilson
coefficients at the scale $\mu = m_b$ in the \LL\ approximation are then given by
\beq
C_i(m_b) = \sum_{j,k}(S^{-1})_{ij}(\eta^{3\lambda_j/23})S_{jk}C_k(M_W) ,
\eeq
where the $\lambda_j$'s in the exponent of $\eta=\alpha_s(M_W)/\alpha_s(m_b)$ are the eigenvalues
of the anomalous dimension matrix over $g^2/16\pi^2$ and the matrix $S$ contains
the corresponding eigenvectors.  The result for the photonic and gluonic magnetic coefficients
are calculated in Ref. \cite{Cho94} and in Ref. \cite{Baren98}, respectively, and the rest of them
related to our analysis can be found in Ref. \cite{Buch96}.\footnote{Although \QCD\ correction
factors in $C^\prime_{1,2}$ are different from those in $C_{1,2}$ in general \cite{Ecker85},
we use an approximation $\alpha_s(M_{W^\prime})\simeq \alpha_s(M_W)$ for simplicity,
which will not change our result.}
Therefore we do not repeat them here, and lead the reader to the original papers.  For 5 flavors,
we have the following numerical values of $C_i(m_b)$ in \LL\ precision using
$\Lambda_{\overline{MS}}$=225 MeV, $m_b$=4.4 GeV, and $m_t$=170 GeV:\footnote{The numbers we
obtained for $C_7^{\gamma(\prime)}$ and $C_8^{G(\prime)}$ are slightly different from those in Ref.
\cite{Baren98} because they used $m_t/m_b$=60.}
\bdpm
C_1^q = -0.308 , \qquad C_1^{q\prime} = C_1^q\zeta_g\lambda^{RR}_q/\lambda^{LL}_q ,
\edpm
\bdpm
C_2^q = 1.144 , \qquad C_2^{q\prime} = C_2^q\zeta_g\lambda^{RR}_q/\lambda^{LL}_q ,
\edpm
\bdpm
C_3 = 0.014 , \quad C_4 = -0.030 , \quad C_5 = 0.009 , \quad C_6 = -0.038 ,
\edpm
\beq \label{Coef-num}
C_7 = 0.045\alpha , \quad C_8 = 0.048\alpha , \quad C_9 = -1.280\alpha ,\quad C_{10}= 0.328\alpha ,
\eeq
\bdpm
C_7^\gamma = -0.317-0.546A^{tb} , \qquad C_7^{\gamma\prime} = -0.546A^{ts\ast} ,
\edpm
\bdpm
C_8^G = -0.150-0.241A^{tb} , \qquad C_8^{G\prime} = -0.241A^{ts\ast}.
\edpm
Note that $C'_3 - C'_{10}$ are negligible comparing to $C_7^{\gamma\prime}$ and $C_8^{G\prime}$
whereas $C'_{1,2}$ are not.  We will show that $C'_{1,2}$ are important to the absorptive
parts in penguin-dominated $B$ decays in the next section.

\begin{figure}[!hbt]
\centering%
  \subfigure[]{\label{O12} %
    \includegraphics[width=4cm]{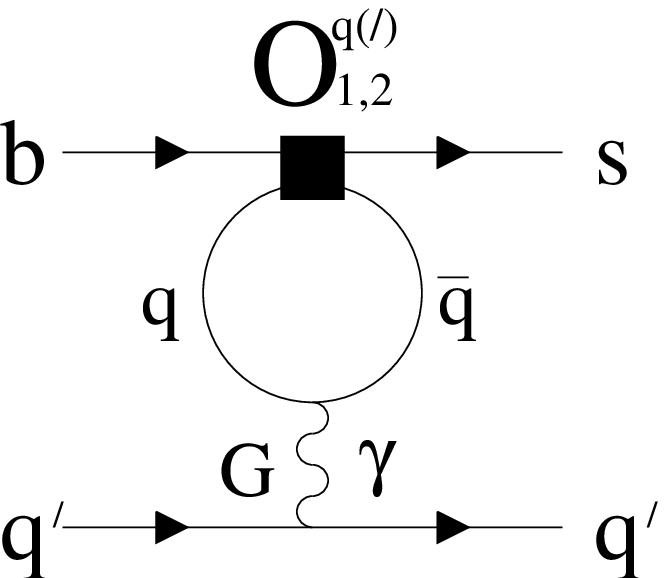}}\hfill
  \subfigure[]{\label{O310} %
    \includegraphics[width=3.5cm]{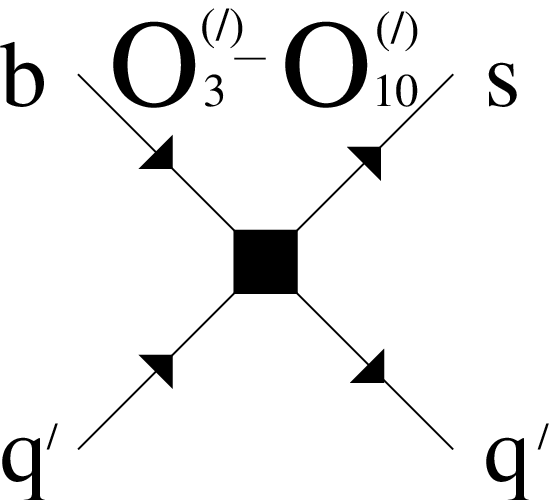}}\hfill
  \subfigure[]{\label{O78} %
    \includegraphics[width=4cm]{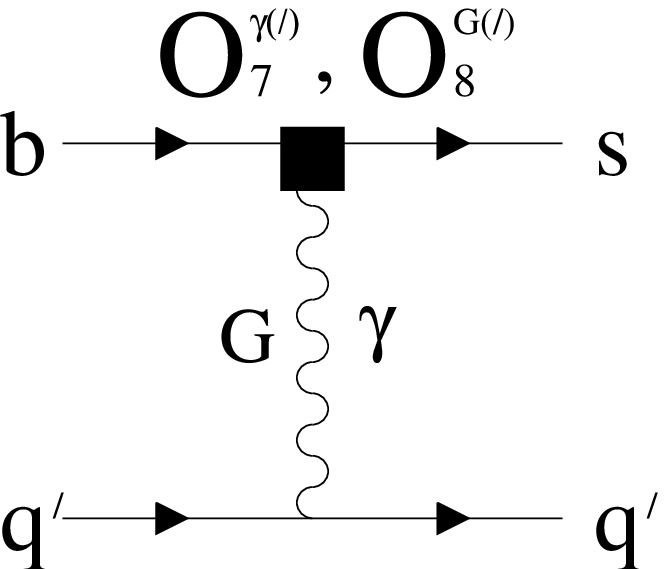}}
\caption{Diagrams for penguin-induced $b \rightarrow s\bar{q^\prime}q^\prime$ decays.}
\label{POP}
\end{figure}

\section{CP violating asymmetries}

\subsection{Charged B meson decays}

  For charged B meson decays, the non-zero \CP\ violating asymmetry defined as
\beq
A_{CP} = \frac{\Gamma(B^+ \rightarrow f^+) - \Gamma(B^- \rightarrow f^-)}
              {\Gamma(B^+ \rightarrow f^+) + \Gamma(B^- \rightarrow f^-)}
\eeq
originates from the superposition of \CP-odd(violating) phases introduced by CKM matrix elements
and \CP-even(conserving) phases arising from the absorptive part of the amplitudes.  Since we
have obtained the relevant effective Hamiltonian in Sec. 2, it is quite straightforward
to calculate the partial decay rates and \CP\ asymmetries in $b \rightarrow s\bar{s}s$ decays.
These decays are governed by three different types of penguin diagrams shown in Fig. \ref{POP}.
The absorptive part of the amplitudes arises at $O(\alpha_s)$ from the one-loop penguin diagrams
with insertions of the operators $O^{(\prime)}_{1,2}$ shown in Fig. \ref{O12}.
The detailed calculation of the one-loop penguin matrix element of the operators $O_{1,2}$
in the \SM\ is in Ref. \cite{Fleischer93} so that we can be very brief.
The renormalized matrix elements of the operators $O^{(\prime)}_{1,2}$ in the \LL\ approximation
are given by
\beqr \label{Otree}
<O^{q(\prime)}_1>^{\mathrm{peng}} &=& \frac{\alpha}{3\pi}
  \mathcal{I}(m_q,k,m_b)<P_\gamma^{(\prime)}> , \cr
<O^{q(\prime)}_2>^{\mathrm{peng}} &=& \frac{\alpha_s(m_b)}{8\pi}
  \mathcal{I}(m_q,k,m_b)\Big(<P_G^{(\prime)}>
  + \frac{8}{9}\frac{\alpha}{\alpha_s(m_b)}<P_\gamma^{(\prime)}>\Big),
\eeqr
where
\beqr
P_G^{(\prime)} &=& O_4^{(\prime)}+O_6^{(\prime)}-\frac{1}{N_c}(O_3^{(\prime)}+O_5^{(\prime)}), \cr
P_\gamma^{(\prime)} &=& O_7^{(\prime)}+O_9^{(\prime)} \quad (N_c = 3),
\eeqr
and
\beq \label{functionI}
\mathcal{I}(m,k,\mu) = 4\int_0^1 dx x(1-x)\ln \Big[\frac{m^2 - k^2x(1-x)}{\mu^2}\Big],
\eeq
and where $k$ is the momentum transferred by the gluon to the ($s,\bar{s}$) pair.
As one can see from Eq. \ref{functionI}, different \CP -even phases arise from the imaginary
parts of the functions $\mathcal{I}(m_u,k,\mu)$ and $\mathcal{I}(m_c,k,\mu)$.
On the other hand, the penguin operators $O_3 - O_{10}$ contribute to only the dispersive parts
of the amplitudes and give tree-level penguin transition amplitudes shown in Fig. \ref{O310}.
Also, as shown in Fig. \ref{O78}, we should include the tree-level diagram associated with
the magnetic operators $O_7^{\gamma(\prime)}$ and $O_8^{G(\prime)}$ to the dispersive part
of the amplitude.  Using the factorization approximation \cite{Bauer85},
we use the following parametrization:
\beqr
<O_7^{\gamma(\prime)}>^{\mathrm{peng}}
  &=& -\frac{\alpha}{3\pi}\frac{m_b^2}{k^2}<P_\gamma^{(\prime)}> , \cr
<O_8^{G(\prime)}>^{\mathrm{peng}}
  &=& -\frac{\alpha_s}{4\pi}\frac{m_b^2}{k^2}<P_G^{(\prime)}> .
\eeqr
Here $k^2$ is expected to be typically in the range $m_b^2/4\leq k^2\leq m_b^2/2$ \cite{Desh90}.
We will use $k^2=m_b^2/2$ for our numerical analysis.

  Now we are ready to consider $B^\pm \rightarrow \phi K^\pm$ decays explicitly.
Since the axial-vector parts of the operators do not contribute to the transition amplitudes
in these decays we can simply use $<O_i>=<O_i^\prime>$ with the help of the vacuum-insertion
method \cite{Gaillard74}.
Combining all operators, we obtain the following transition amplitude using the unitarity
relation $\sum_{q=u,c,t}\lambda_q=0$:
\beqr \label{asymfn}
\mathcal{A}(B^-\rightarrow \phi K^-) &=& \frac{G_F}{\sqrt{2}}\sum_{q=u,c}\lambda_q^{LL}
  \Bigg[\frac{\alpha_s(m_b)}{9\pi}\Big\{C_2^q(m_b) \cr
&&-\frac{7}{6}\frac{\alpha}{\alpha_s(m_b)}(3C_1^q(m_b)+C_2^q(m_b))\Big\}\mathcal{I}(m_q,k,m_b)\cr
&&-\frac{\alpha_s(m_b)}{9\pi}\Big\{4C_8^G(m_b)-7\frac{\alpha}{\alpha_s(m_b)}C_7^\gamma(m_b)\Big\}\\
&&+\frac{4}{3}(C_3(m_b)+C_4(m_b))+C_5(m_b)+\frac{1}{3}C_6(m_b) \cr
&&-\frac{1}{2}C_7(m_b)-\frac{1}{6}C_8(m_b)-\frac{2}{3}(C_9(m_b)+C_{10}(m_b))\Bigg] \cr
&&\times  X^{(B^-K^-,\phi)} + (C_i\rightarrow C'_i), \nonumber
\eeqr
where $X^{(B^-K^-,\phi)}\equiv <\phi |\bar{s}\gamma_\mu s|0><K^-|\bar{s}\gamma^\mu b|B^->$.
The amplitude $\mathcal{A}(B^+\rightarrow \phi K^+)$ is simply obtained from
$\mathcal{A}(B^-\rightarrow \phi K^-)$ by replacing
$\lambda_q^{LL}\rightarrow \lambda_q^{LL\ast}$ and $C_i^{(\prime)}\rightarrow C_i^{(\prime)\ast}$.
In the \SM, non-zero \CP\ asymmetry arises from the superposition of the \CP
-odd phase $\gamma$ in $V_{ub}^L$ and the different \CP -even phases arising from the function
$\mathcal{I}(m_q,k,m_b)$ due to the mass difference between $c$- and $u$-quark.
The resulting \CP\ asymmetry is known to be very small $\sim O(10^{-2})$ \cite{Fleischer93,Ali98}
because the magnitude of the absorptive part is much smaller than that of the dispersive part.
Using the numbers in Eq. \ref{Coef-num}, $m_c$=1.3 GeV, and Arg$[V_{ub}^L]=-59^\circ$,
we can estimate the \SM\ value of \CP\ asymmetry:
\beq
A_{CP}^{SM}(B^\pm\rightarrow \phi K^\pm)\simeq 7.3\times 10^{-3} .
\label{CPSM}
\eeq
If the model has manifest left-right symmetry, the $W_R$ mass has a stringent bound
$M_{W_R}\geq$1.6 TeV \cite{Beall82}, and its contribution to the decay amplitude is very small
so that \CP\ asymmetry in the manifest \LRM\ should be very small as well.
Since this value is small and our purpose is to estimate the possible large right-handed
current contribution, we take a limit $\mathcal{I}(m_c,k,\mu)=\mathcal{I}(m_u,k,\mu)$
in order to get around the uncertainty of $V_{ub}^L$ obtained under the \SM\ framework
and clearly see the right-handed current contribution.
Then we can express $\mathcal{A}(B^-\rightarrow \phi K^-)$ in terms of new parameters $\zeta_g$, $\xi_g$,
and $\theta_R$ for two types of $V^R$ in Eq. (\ref{VR}) in the \LRM\ using the unitarity relation
$\sum_{q=u,c,t}\lambda_q=0$ and the numbers in Eq. \ref{Coef-num} again as follows:
\beqr
\mathcal{A}(B^-\rightarrow \phi K^-)_I &\simeq& -\frac{G_F}{\sqrt{2}}\Big\{
 -2.87e^{i\varphi_1} + 23.1e^{i\varphi_2}\zeta_g c_Rs_Re^{i(\alpha_4 - \alpha_3)} \cr
 &&+ 10.1\xi_g(c_Re^{i\alpha_4} - 25s_Re^{i\alpha_3}) \Big\}\times 10^{-3}X^{(B^-K^-,\phi)} , \\
\mathcal{A}(B^-\rightarrow \phi K^-)_{II} &\simeq& -\frac{G_F}{\sqrt{2}}\Big\{
 -2.87e^{i\varphi_1} + 10.1\xi_g c_Re^{i\alpha_4} \Big\}\times 10^{-3}X^{(B^-K^-,\phi)} ,\nonumber
\label{BmtoKm}
\eeqr
where ($\varphi_1, \varphi_2$)=($-14.9^\circ , -53.1^\circ$) are \CP -even phases.
As stated earlier, one can clearly see here that the $\zeta_g$ term coming from the coefficients
$C'_{1,2}$ is not negligible in case of $V^R_I$.
Likewise, the transition amplitude in $B^-\rightarrow \phi K^{\ast -}$ decays can be easily
obtained by using $<O_i>=-<O_i^\prime>$ because $K^{\ast -}$ is a vector particle:
\beqr
\mathcal{A}(B^-\rightarrow \phi K^{\ast -})_I &\simeq& -\frac{G_F}{\sqrt{2}}\Big\{2.87e^{i\varphi_1}
+ 23.1e^{i\varphi_2}\zeta_g c_Rs_Re^{i(\alpha_4 - \alpha_3)} \cr
&&+ 10.1\xi_g(-c_Re^{i\alpha_4} - 25s_Re^{i\alpha_3})\Big\}\times 10^{-3}X^{(B^-K^{\ast -},\phi)},
\\
\mathcal{A}(B^-\rightarrow \phi K^{\ast -})_{II} &\simeq&
 -\frac{G_F}{\sqrt{2}}\Big\{2.87e^{i\varphi_1}
- 10.1\xi_g c_Re^{i\alpha_4} \Big\}\times 10^{-3}X^{(B^-K^{\ast -},\phi)} , \nonumber
\label{BmtoKmx}
\eeqr
where $X^{(B^-K^{\ast -},\phi)}\equiv
<\phi |\bar{s}\gamma_\mu s|0><K^{\ast-}|\bar{s}\gamma^\mu\gamma_5 b|B^->$.  Although the \CP\
asymmetry in $B^-\rightarrow \phi K^-$ decays should be the same as that in
$B^-\rightarrow \phi K^{\ast -}$ decays in the \SM , they can be different in \LRM\ so that
the measured difference of \CP\ asymmetries between them may give the size of the \NP\ effects.

\begin{figure}[!hbt]
\centering%
  \subfigure[$B^\pm\rightarrow \phi K^\pm$ decays]{\label{K1a} %
    \includegraphics[width=7cm]{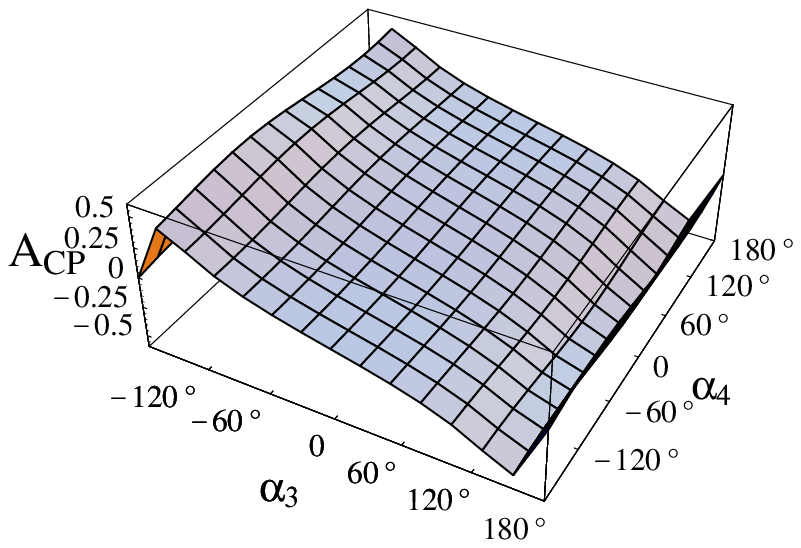}} \quad
  \subfigure[$B^\pm\rightarrow \phi K^{\ast\pm}$ decays]{\label{K1b} %
    \includegraphics[width=7cm]{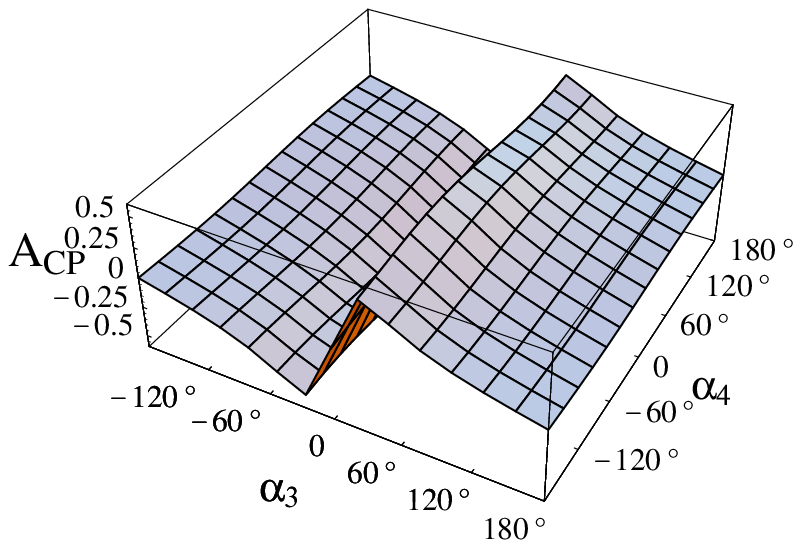}}
\caption{Behavior of $A_{CP}$ as $\alpha_{3,4}$ are varied in the case of $V^R_I$.}
\label{K1}
\end{figure}

\begin{figure}[!hbt]
\centering%
    \includegraphics[width=7cm]{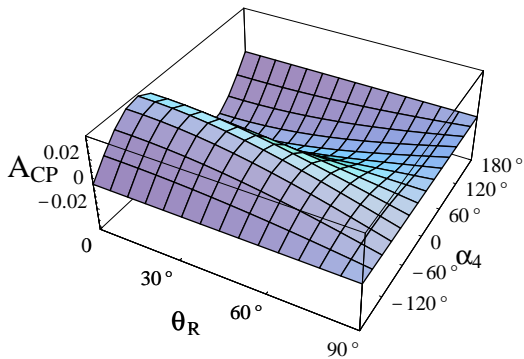}
\caption{Behavior of $A_{CP}(B^\pm\rightarrow \phi K^{(\ast)\pm})$
as $\theta_R$ and $\alpha_4$ are varied in the case of $V^R_{II}$.}
\label{K2}
\end{figure}

  The current data on the \CP\ asymmetries in $B^-\rightarrow \phi K^-$ and
$B^-\rightarrow \phi K^{\ast -}$ decays are \cite{Aubert02}:
\beqr
A_{CP}^{\mathrm{expt}}(B^\pm\rightarrow \phi K^\pm) &=& 0.05 \pm 0.20 \pm 0.03 , \cr
A_{CP}^{\mathrm{expt}}(B^\pm\rightarrow \phi K^{\ast\pm}) &=& 0.43^{\ + \ 0.36}_{\ - \ 0.30}
\pm 0.06 .
\eeqr
The \SM\ value in Eq. (\ref{CPSM}) lies in the range of
$A_{CP}^{\mathrm{expt}}(B^\pm\rightarrow \phi K^\pm)$, but a little off the range of
$A_{CP}^{\mathrm{expt}}(B^\pm\rightarrow \phi K^{\ast\pm})$. In order to explicitly compare
these values with the theoretical estimates in the \LRM , we first plot
$A_{CP}(B^\pm\rightarrow \phi K^\pm)$ and $A_{CP}(B^\pm\rightarrow \phi K^{\ast\pm})$
in the case of $V_I^R$ in Fig. \ref{K1} for the typical values $\zeta_g$=0.01, $\xi_g$=0.008, and
$\theta_R=70^\circ$ as $\alpha_{3,4}$ are varied.  In the figure, \CP\ asymmetry is
drastically changing by varying $\alpha_3$, and this behavior
holds for other values of $\zeta_g$, $\xi_g$, and $\theta_R$.  For the given inputs,
$A_{CP}(B^\pm\rightarrow \phi K^\pm)$ and $A_{CP}(B^\pm\rightarrow \phi K^{\ast\pm})$ can be
different by about 0.5.  In the case of $V^R_{II}$, one can see from Eqs. (\ref{BmtoKm}),
(\ref{BmtoKmx}) that $A_{CP}(B^\pm\rightarrow \phi K^\pm)$ =
$A_{CP}(B^\pm\rightarrow \phi K^{\ast\pm})$ because it has no dependance of $\zeta_g$ and
$\alpha_3$ unlike the previous case.  In Fig. \ref{K2},  we fix $\xi_g$=0.01,
and evaluate \CP\ asymmetry by varying $\theta_R$ and $\alpha_4$.  It shows that \CP\ asymmetry
is very small with a small parameter $\xi_g$.
Therefore, if we observe large \CP\ asymmetry or any difference between
$A_{CP}(B^\pm\rightarrow \phi K^\pm)$ and $A_{CP}(B^\pm\rightarrow \phi K^{\ast\pm})$,
the second type of mass mixing matrix $V_{II}^R$ is disfavored.

\subsection{Neutral B meson decays}

In the case of the neutral $B$ meson decays into \CP\ self-conjugate final states $f$,
mixing induced \CP\ asymmetry can be expressed by the parametrization invariant quantity
$\lambda$ defined by \cite{Review}
\beq
\lambda \equiv \eta_f \left( \frac{q}{p}\right)_B \frac{\mathcal{A}(\bar{B^0}
  \rightarrow \bar{f})}{\mathcal{A}(B^0 \rightarrow f )} , \qquad \left(\frac{q}{p}\right)_B
  \simeq \frac{M^*_{12}}{|M_{12}|} ,
\eeq
where $\eta_f$=1(-1) for a \CP -even(odd) final state $f$ and $M_{12}$ is the dispersive
part of the $B\bar{B}$ mixing matrix element.  The \CP\ angle $\beta$ mentioned earlier
is simply the imaginary part of $\lambda$ in $B \rightarrow J/\psi K_S$ decays in the \SM :
\beq
\sin 2\beta =  \textrm{Im} \lambda (B \rightarrow J/\psi K_S )
  \simeq \textrm{Im} \lambda (B \rightarrow \phi K_S ) .
\eeq
In the general \LRM , $M_{12}$ can be written as
\beq \label{massmixing}
M_{12} = M_{12}^{SM} + M_{12}^{LR} = M_{12}^{SM}\biggl\{ 1 + r_{LR} \biggr\} ,
\eeq
where
\beq
r_{LR} \equiv \frac{M_{12}^{LR}}{M_{12}^{SM}}
  = \frac{<\bar{B^0}|H_{eff}^{LR}|B^0>}{<\bar{B^0}|H_{eff}^{SM}|B^0>} ,
\eeq
with the effective Hamiltonian
$H^{B\bar{B}}_{eff} = H^{SM}_{eff} + H^{LR}_{eff}$ in the $B\bar{B}$ system.
Considering the two types of the quark mixing matrices in Eq. (\ref{VR}),
the effective Hamiltonians in the $B\bar{B}$ system are given by
\beqr
H^{SM}_{eff} &=& \frac{G_F^2M_W^2}{4\pi^2}(\lambda_t^{LL})^2S(x^2_t)
                (\bar{d_L}\gamma_\mu b_L)^2 , \label{HSMeff}\\
H^{LR}_{eff} &=& \frac{G_F^2M_W^2}{2\pi^2} \biggl[
                  \{\lambda_c^{LR} \lambda_t^{RL}x_cx_t\zeta_g A_1(x_t^2,\zeta)
                 + \lambda_t^{LR} \lambda_t^{RL}x_t^2\zeta_g A_2(x_t^2,\zeta)\}
                   (\bar{d_L}b_R)(\bar{d_R}b_L) \cr
  && + \lambda_t^{LL} \lambda_t^{RL}x_b\xi_g
                    \{x_t^3A_3(x_t^2)(\bar{d_L}\gamma_\mu b_L)(\bar{d_R}\gamma_\mu b_R)
     + x_tA_4(x_t^2)(\bar{d_L}b_R)(\bar{d_R}b_L)\}\biggr] ,
\label{HLReff}
\eeqr
where $S(x)$ is the usual Inami-Lim function and $A_i$ can be found in Ref. \cite{Nam02}.
If we consider \QCD\ effect in $B\bar{B}$ mixing, the correction factors should be
included in the functions $S$ and $A_i$.
However, there are many uncertainties such as hadronic matrix elements and new parameters
in the \LRM\ to prevent us from the precision analysis at this stage, and the \QCD\
corrections to $B\bar{B}$ mixing are not big enough to change our numerical estimate.
Therefore we will ignore the \QCD\ corrections to $B\bar{B}$ mixing for simplicity.
In the case of $V^R_I$, there is no significant
contribution of $H^{LR}_{eff}$ to $B\bar{B}$ mixing, so that $M_{12} = M_{12}^{SM}$
because $\lambda_t^{RL}\simeq 0$.
In the case of $V^R_{II}$,
using $m_c$=1.3 GeV, $m_b$=4.4 GeV, $m_t$=170 GeV, and $|V^L_{cd}|\approx 0.224$,
and adopting the parametrization of the hadronic matrix elements of the operators given
in Ref. \cite{Nam02},
one can express $r_{LR}$ in terms of the mixing angle and phases in Eq. (\ref{VR}) as
\beqr \label{rLR}
r_{LR} &\approx& \textit{l} \biggl\{ 17.3 \textit{l} \biggl(
 \frac{1 - \zeta_g - (4.92 - 19.7\zeta_g)\ln(1/\zeta_g) }{ 1 - 5.47\zeta_g }\biggr)
 \zeta_g s_R^2e^{i\delta_1} \cr
 &-& 796 \biggl( \frac{ 1 - 5.02\zeta_g - (0.498 - 1.99\zeta_g )\ln(1/\zeta_g) }
         {1 - 9.94\zeta_g + 28.9\zeta_g^2}\biggr) \zeta_g s_Rc_Re^{i\delta_2} \
         - \ 8.93\xi_g s_Re^{i\delta_3} \biggr\} ,
\eeqr
where $\textit{l} = 0.008/|V^L_{td}|$, $\delta_1 = -2\beta + \alpha_2 - \alpha_3$,
$\delta_2 = -\beta - \alpha_3 + \alpha_4$, $\delta_3 = -\beta - \alpha_3$.
Since $B \rightarrow J/\psi K_S$ decay is governed by the tree-level amplitude,
the transition amplitude is given by
\beqr
\mathcal{A}(B \rightarrow J/\psi K_s)_I &\simeq& \frac{G_F}{\sqrt{2}}\lambda_c^{LL}
 \Big\{1 + 25(c_Rs_R\zeta_ge^{-i(\alpha_2-\alpha_1)} - 2s_R\xi_ge^{-i\alpha_2}) \Big\}
  X^{(B K_s,J/\psi)} , \cr
\mathcal{A}(B \rightarrow J/\psi K_s)_{II} &\simeq& \frac{G_F}{\sqrt{2}}\lambda_c^{LL}
 \Big\{1 -50s_R\xi_ge^{-i\alpha_2} \Big\}
  X^{(B K_s,J/\psi)} ,
\eeqr
where $X^{(B K_s,J/\psi)}\equiv
<J/\psi |\bar{c}\gamma_\mu c|0><K_s|\bar{s}\gamma^\mu b|B^\circ >$, and we ignored
the $K\bar{K}$ mixing.  The transition amplitude in $B \rightarrow \phi K_S$ decays
can be simply obtained from Eq. (\ref{BmtoKm}) by replacing the hadronic matrix element
$X^{(B^- K^- , \phi)} \rightarrow X^{(B K_s , \phi)}$.

\begin{figure}[!hbt]
\centering%
  \subfigure[]{\label{JP1a} %
    \includegraphics[width=7cm]{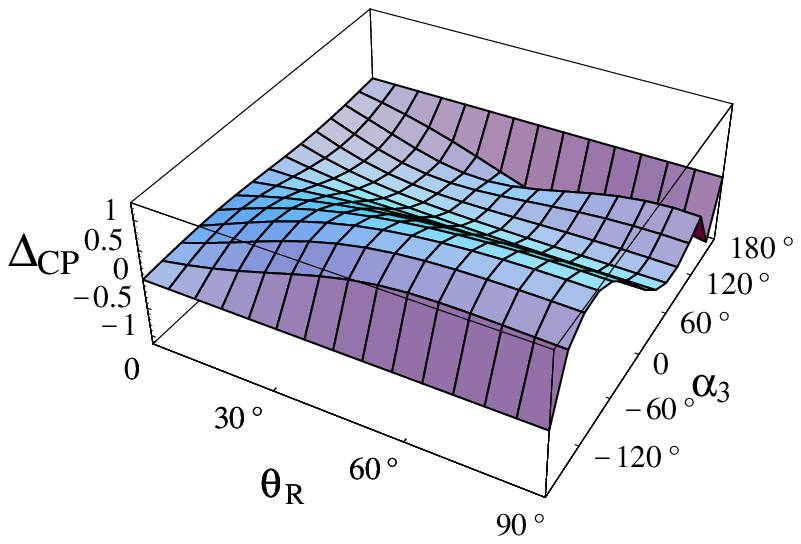}}\quad
  \subfigure[]{\label{JP1b} %
    \includegraphics[width=7cm]{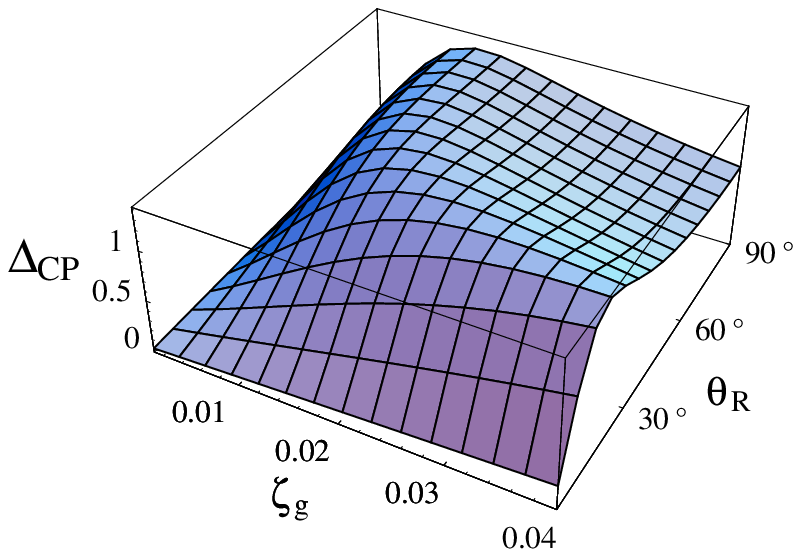}}
\caption{Behavior of the \CP\ asymmetry difference $\Delta_{CP}$ between
$B \rightarrow J/\psi K_S$ and $B \rightarrow \phi K_S$ decays in the case of $V_I^R$. }
\end{figure}

\begin{figure}[!hbt]
\centering%
  \includegraphics[width=6.5cm]{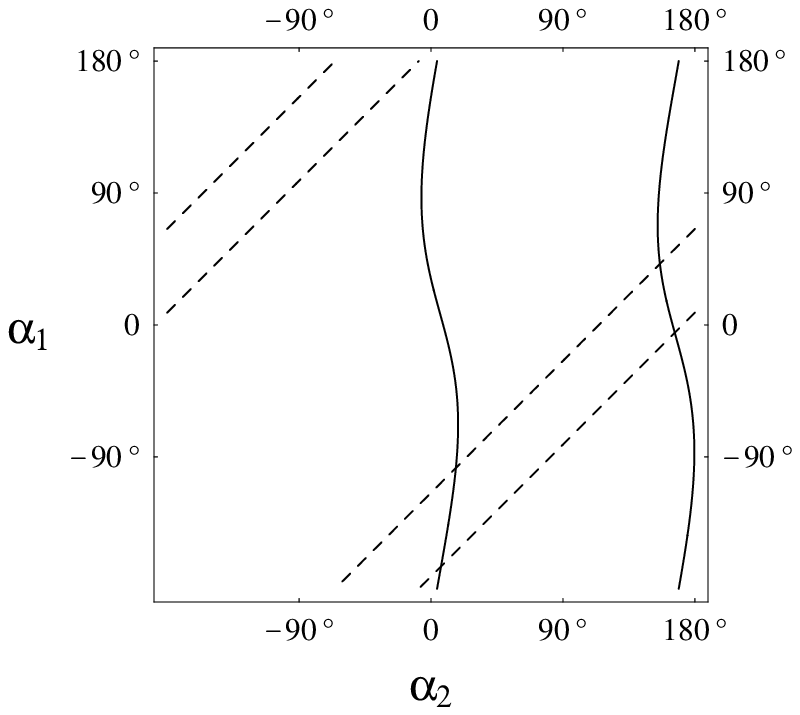}
\caption{Contour plot corresponding to Im$\lambda (B \rightarrow J/\psi K_S) = 0.73$
(solid line) and Im$\lambda (B \rightarrow \phi K_S) = -0.39$ (dashed line)
for sin2$\beta = 0.64$ in the case of $V_I^R$.}
\label{JP1c}
\end{figure}

\begin{figure}[!hbt]
\centering%
  \subfigure[]{\label{JP2a} %
    \includegraphics[width=7cm]{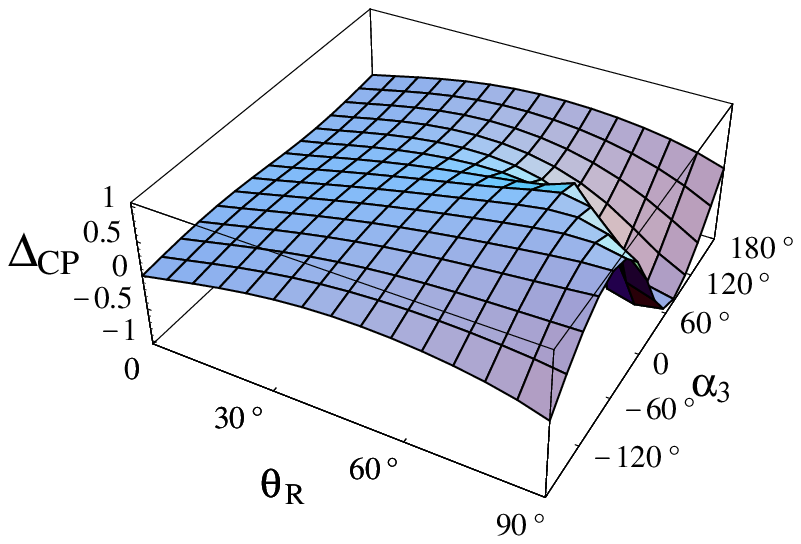}}\quad
  \subfigure[]{\label{JP2b} %
    \includegraphics[width=7cm]{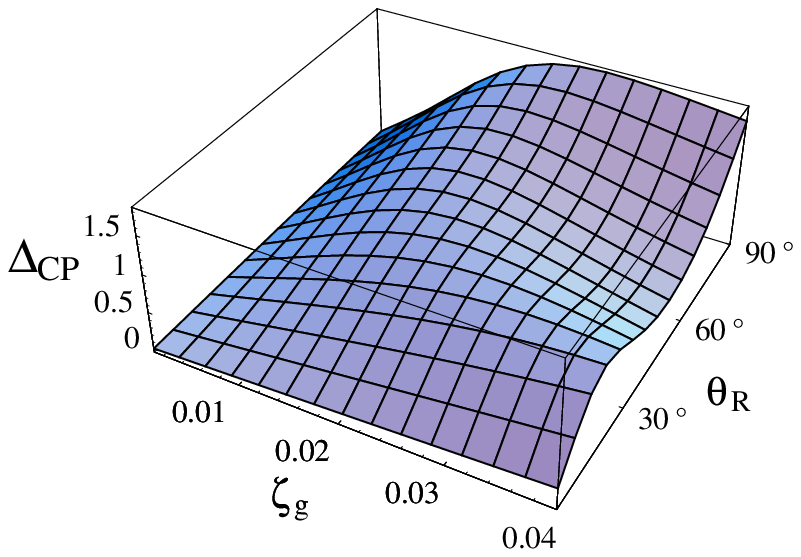}}
\caption{Behavior of the \CP\ asymmetry difference $\Delta_{CP}$ between
$B \rightarrow J/\psi K_S$ and $B \rightarrow \phi K_S$ decays in the case of $V_{II}^R$. }
\label{JP2}
\end{figure}

\begin{figure}[!hbt]
\centering%
  \includegraphics[width=6.5cm]{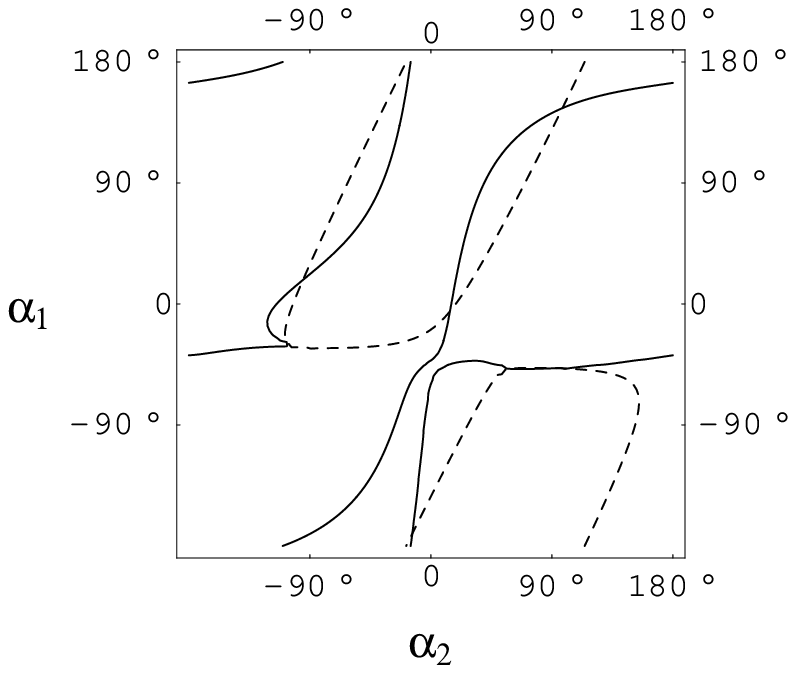}
\caption{Contour plot corresponding to Im$\lambda (B \rightarrow J/\psi K_S) = 0.73$
(solid line) and Im$\lambda (B \rightarrow \phi K_S) = -0.39$ (dashed line)
for sin2$\beta = 0.64$ in the case of $V_{II}^R$.}
\label{JP2c}
\end{figure}

  For illustration of the possible effect of the new interaction on the mixing induced
\CP\ asymmetry, we assume that $\beta = 20^\circ$ and $\textit{l} = 1$, and show that
the region of parameters $\alpha_i$ where Im$\lambda (B \rightarrow J/\psi K_S) \simeq 0.73$ and
Im$\lambda (B \rightarrow \phi K_S) \simeq -0.39$ since $|\lambda |\approx 1$.  To do so,
we need to find an appropriate set of parameters $\zeta_g$, $\xi_g$, and $\theta_R$
yielding a large difference $\Delta_{CP} \equiv \textrm{Im}\lambda (B \rightarrow J/\psi K_S) -
\textrm{Im}\lambda (B \rightarrow \phi K_S)$.
First, we evaluate $\Delta_{CP}$ in the case of $V^R_I$ for
$\zeta_g = \xi_g = 0.01$, $\alpha_{1,2} = 0$ by varying $\theta_R$ and $\alpha_3$
in Fig. \ref{JP1a}.  In the figure, $\Delta_{CP}$ becomes maximal near
$\alpha_3 \sim -120^\circ$ and increases as $\theta_R$ increases,
and this behavior holds for other values of fixed parameters.
Since we assumed that $\Delta_{CP}$ is larger than 1, we fix $\alpha_3 = -120^\circ$,
and evaluate $\Delta_{CP}$ in Fig. \ref{JP1b} for $\alpha_{1,2} = 0$ and
$\xi_g = \zeta_g$ by varying $\theta_R$ and $\zeta_g$.  One can see from the figure that
$\Delta_{CP}$ approaches 1 for $\zeta_g \gtrsim 0.01$ and $\theta_R \gtrsim 10^\circ$,
and its variation is small.  After repeating this analysis, we get a probable set
of parameter values $\zeta_g = 0.01$, $\xi_g = 0.008$, $\theta_R = 70^\circ$,
and $\alpha_3 = -120^\circ$.  Using these values, we plot the contours corresponding to
Im$\lambda (B \rightarrow J/\psi K_S) = 0.73$ and
Im$\lambda (B \rightarrow \phi K_S) = -0.39$ in the parameter space of $\alpha_{1,2}$
in Fig. \ref{JP1c}.  Therefore, as a result from the obtained figures,
the manifest or pseudomanifest \LRM\ is disfavored under the given assumption.
In a similar way to the case of $V^R_I$, the results of the analysis of the mixing induced
\CP\ asymmetries in the case of $V^R_{II}$ are represented in Fig. \ref{JP2} and Fig. \ref{JP2c}.

\section{Conclusions}

  In this paper, we studied \CP\ asymmetries in penguin-induced $b \rightarrow s\bar{s}s$ decays
in the general \LRM . Without imposing manifest or pseudomanifest left-right symmetry, one has
two types of mass mixing matrix $V^R$ with which the right-handed current contributions
to $B\bar{B}$ mixing and \CP\ asymmetry can be sizable even in the decays such as
$B^\pm \rightarrow \phi K^{(\ast)\pm}$ decays where the \SM\ contribution to
\CP\ asymmetry is very small.  Using the effective Hamiltonian approach, we evaluate the sizes
of the \NP\ contributions to \CP\ asymmetries in $B^\pm \rightarrow \phi K^{(\ast)\pm}$ decays,
and show that $V^R_I$ is more probable than $V^R_{II}$ if \CP\ asymmetries in those decays are
large or different from each other.  Similar argument can be made in mixing induced $B$ decays
such as $B \rightarrow J/\psi K_s$ and $B \rightarrow \phi K_s$ decays.  Although \SM\ predicts
that the \CP\ asymmetry in $B \rightarrow J/\psi K_s$ decays should be very close to that in
$B \rightarrow \phi K_s$ decays, the present experiments show a large discrepancy between them.
Based on these preliminary experimental results, we find that the manifest or pseudomanifest
\LRM\ is disfavored, and the bounds of the new parameters are restricted as shown in Figs. 4-7.
Furthermore, this result may affect the sizes of \CP\ asymmetries in other decays.
For instance, one can see from Fig. \ref{K1} and Fig. \ref{K2} that the contributions of
the obtained parameter sets from Fig. \ref{JP1c} and Fig. \ref{JP2c} under the given assumption
reduces the size of \CP\ asymmetries in $B^\pm \rightarrow \phi K^{\ast\pm}$ decays.
In this way, \CP\ asymmetries in other mixing induced decays such as $B \rightarrow \phi K^\ast$
can be estimated systematically, and all of these analysis of possible \NP\ contributions
can be tested once the experimental results are confirmed.

\section{Acknowledgements}
The author would like to thank M.B. Voloshin for careful reading of the manuscript and
his valuable comments.
This work is supported in part by the DOE grant DE-FG02-94ER40823.

\newpage

\end{document}